\documentstyle[aps,prd,amsmath,amssymb,amscd,epsfig,subfigure,tabularx,preprint]{revtex}

\def\beq{\begin{eqnarray}}
\def\eeq{\end{eqnarray}}
\def\bsp{\begin{split}}
\def\esp{\end{split}}
\def\lra{\longrightarrow}
\def\ra{\rightarrow}
\def\inl{\left\langle}
\def\inr{\right\rangle}

\newcommand{\mb}[1]{{\mathbb #1}}

\tightenlines
\begin{document}
\title{\bf Gravitational Entropy and Quantum Cosmology}
\author{\O yvind Gr\o n$^{1,2}$\footnote{e-mail:Oyvind.Gron@iu.hioslo.no} and 
Sigbj\o rn Hervik$^1$\footnote{e-mail:sigbjorn.hervik@fys.uio.no}\\
\vspace{1cm}
$^1$Department of Physics, University of Oslo\\
P.O.Box 1048 Blindern\\
N-0316 Oslo, Norway\\
\vspace{.5cm}
$^2$Oslo College, Faculty of Engineering \\
Cort Adelers gt. 30\\
N-0254 Oslo, Norway}
\maketitle

\begin{abstract}

We investigate the evolution of different measures of ``Gravitational Entropy''
 in Bianchi type I and Lema\^itre-Tolman universe models.

A new quantity behaving in accordance with the second law of thermodynamics is 
introduced. We then go on and investigate whether a quantum calculation of 
initial conditions for the universe based upon the Wheeler-DeWitt equation 
supports Penrose's {\it Weyl Curvature Conjecture}, according to which the 
Ricci part of the curvature dominates over the Weyl part at the initial 
singularity of the universe. The theory is applied to the Bianchi type I 
universe models with dust and a cosmological constant and to the 
Lema\^itre-Tolman universe models. We investigate two different versions of the
 conjecture. First we investigate a local version which fails to support the 
conjecture. Thereafter we construct a non-local entity which shows more 
promising behaviour concerning the conjecture. 
\end{abstract}

\section{Introduction}

The physics of the arrow of time in our universe seems somehow to involve 
gravitation\cite{davies74}. The origin of this arrow may be searched for in 
the initial state and early evolution of the universe \cite{davies83}. 

The isotropy of the cosmic background radiation shows that the universe was to
 a very high degree in a state of thermodynamic equilibrium already 300.000 
years after the Big Bang. At the present time there are great thermal 
differences between matter in the stars and that in the interstellar clouds. 
Matter has developed away from a state of thermal equilibrium instead of 
towards it, in contradiction to what one could expect from the second law of 
thermodynamics. This is clearly due to gravity, which increases density 
fluctuations and hence increases the cosmic temperature differences.

One would like to incorporate this tendency of gravity to produce 
inhomogeneities into a generalized second law of thermodynamics. Then one needs
 a quantity representing the entropy of a gravitational field. This quantity 
should vanish in the case of a homogeneous field and obtain a maximal value 
given by the Bekenstein-Hawking \cite{bekenstein,hawking} entropy of a black 
hole, for a gravitational equilibrium configuration in the form of the field 
of a black hole.

One suggestion in this connection was Penrose's formulation of what is called 
the {\it Weyl curvature conjecture}(WCC) \cite{penrose}. Wainwright and 
Anderson \cite{wa} express this conjecture in terms of the ratio of the Weyl 
and the Ricci curvature invariants,
\beq
P^2=\frac{C_{\alpha \beta\gamma\delta}C^{\alpha \beta\gamma\delta}}{R_{\mu\nu}
R^{\mu\nu}}
\eeq
According to the conjecture $P^2$ vanishes at the initial singularity of the 
universe. The physical content of the conjecture is that the initial state of 
the universe is homogeneous and isotropic. As pointed out by Rothman and 
Anninos \cite{ra,rothman} the entities $P^2$ and 
$C_{\alpha \beta\gamma\delta}C^{\alpha \beta\gamma\delta}$ are \emph{local} 
entities in contrast to what we usually think of entropy. In this paper we will
 investigate various entities  in the Bianchi type I and the Lema\^itre-Tolman
 models. We will give a careful investigation of both the entities 
$C_{\alpha \beta\gamma\delta}C^{\alpha \beta\gamma\delta}$ and $P^2$, and show
 that both of these fail to describe a behaviour according to the WCC. We will
 therefore introduce a non-local entity which shows a more promising behaviour
 concerning the WCC. This entity is also constructed in terms of the Weyl 
tensor, and it has therefore a direct connection with the Weyl curvature 
tensor but in a non-local form. The local version of the WCC is a much 
stronger restriction on the cosmological models and it is not strange that it 
fails for a general model.

The vanishing of the Weyl curvature tensor at the initial singularity is a 
very special initial condition. Due to the evolution during the inflationary 
era, however, this condition may be relaxed. A much larger variety of initial 
conditions are consistent with the homogeneity of the universe, as observed in
 the cosmic background radiation, in inflationary universe models than in 
models without inflation. As we shall see from the quantum calculations, an 
inhomogeneous universe is more likely to be spontaneously created than a 
homogeneous one. A universe with a large cosmological constant is also more 
likely to be created than a universe with a small cosmological constant. 

In sections II and III, respectively, we present a classical cosmological 
investigation of  $C_{\alpha \beta\gamma\delta}C^{\alpha \beta\gamma\delta}$ 
and $P^2$ for the Bianchi type I and Lema\^itre-Tolman universe models. Both 
of these entities fail to have the right entropic behavior. We therefore 
propose in section IV an entity which is non-local and shows a much better 
behaviour concerning the WCC. It should be noted however that this entity is 
by no means the true {\it entropy for the gravitational field} but it does 
have the right entropic behaviour. In section V we give a quantum cosmological
 treatment of the WCC in the stage set by the classical investigations. 

\section{The Weyl Curvature Conjecture for a Bianchi type I model: Local 
version}

Based on the results obtained in an earlier paper where the Bianchi type I 
minisuperspace model was investigated\cite{sigBI}, we will consider the Weyl 
curvature conjecture (WCC) for a Bianchi type I model. Writing 
\beq 
ds^2=-dt^2+e^{2\alpha}\left[e^{2\beta}\right]_{ij}dx^idx^j
\eeq
where 
$\beta=\text{diag}(\beta_++\sqrt{3}\beta_-,\beta_+-\sqrt{3}\beta_-,-2\beta_+)$
 the Weyl curvature invariant and the Ricci square turn out to be:
\begin{align}
\bsp
C_{\alpha \beta\gamma\delta}C^{\alpha \beta\gamma\delta} &=
4\big{[}12(\dot{\beta}_+^2+\dot{\beta}_-^2)^2+3(\ddot{\beta}_+^2+
\ddot{\beta}_-^2)+3\dot{\alpha}^2(\dot{\beta}_+^2+\dot{\beta}_-^2) \\
&-12\dot{\alpha}\dot{\beta}_+(\dot{\beta}_+^2-3\dot{\beta}_-^2)+
3\dot{\alpha}\frac{d}{dt}(\dot{\beta}_+^2+\dot{\beta}_-^2)-
\frac{d}{dt}\dot{\beta}_+(\dot{\beta}_+^2-3\dot{\beta}_-^2)\big{]}
\esp\\
\bsp
R^{\mu \nu}{R_{\mu \nu}}&= 36(\dot{\beta}_+^2+\dot{\beta}_-^2)^2+
90\dot{\alpha}^2(\dot{\beta}_+^2+\dot{\beta}_-^2)+36\dot{\alpha}^4+
12\ddot{\alpha}^2 \\
&-36\dot{\alpha}(\dot{\beta}_+\ddot{\beta}_++\dot{\beta}_-\ddot{\beta}_-)+
6(\ddot{\beta}_+^2+\ddot{\beta}_-^2)-36\ddot{\alpha}(\dot{\beta}_+^2+
\dot{\beta}_-^2)-36\ddot{\alpha}\dot{\alpha}^2
\esp\end{align}
Introducing the volume element $v=e^{3\alpha}$, the Einstein field equations 
yield the  equation for $v$:
\begin{equation}\label{volum}
\dot{v}^2=3\Lambda v^2+3Mv+A^2
\end{equation}
Here are $A$ an anisotropy parameter, $M$ the total mass of the dust and 
$\Lambda$ a cosmological constant. The equations for $\beta_{\pm}$ are 
\beq\label{beta}
\dot{\beta}_{\pm}=\frac{a_{\pm}}{3v}
\eeq
where $a_{\pm}$ are constants satisfying $a_+^2 +a_-^2=A^2$. Thus we can 
define a new angular variable $\gamma$ by $a_+=A\sin (\gamma-\frac{\pi}{6})$ 
and $a_-=A\cos (\gamma-\frac{\pi}{6})$ that characterizes the different 
solutions. \cite{sigBI}

\subsection{The Weyl Curvature Tensor for a classical Bianchi type I model}
In the classical theory we have no factor ordering problems. Inserting the 
classical equations \ref{volum} and \ref{beta} and expressing them in terms of 
the volume element $v=e^{3\alpha}$, we get the Weyl curvature invariant:
\begin{equation}
C^{\alpha \beta \gamma \delta}C_{\alpha \beta \gamma \delta}=
\frac{16}{27}\frac{A^4}{v^4}\left[2 \mp 2\sqrt{1+\frac{3M v}{A^2}+
\frac{3\Lambda v^2}{A^2}}\cos 3\gamma +\frac{3M v}{A^2}+
\frac{3\Lambda v^2}{A^2}\right]
\end{equation}
where the $-$ is for expanding and $+$ is for contracting solutions. 
Except for degenerate cases this curvature invariant diverges as 
$\frac{1}{v^4}$ for small values of $v$. For large values of $v$ (if the 
classical equations allow it) this invariant tends to zero. If we write 
$z=\frac{\dot{v}}{A}=\pm\sqrt{1+\frac{3M v}{A^2}+\frac{3\Lambda v^2}{A^2}}$ we
 can write:
\[
C^{\alpha \beta \gamma \delta}C_{\alpha \beta \gamma \delta}=
\frac{16}{27}\frac{A^4}{v^4}\left(1-2z\cos 3\gamma+z^2\right)
\]
As expected, the Einstein-deSitter case, $A=0$, yields a zero Weyl tensor. The
 Weyl tensor is also zero in a few other cases. Let us from now on assume that
 $A\neq 0$. Then iff\footnote{Iff=If and only if} the Weyl tensor is zero, then
 $\cos{3\gamma}=\pm 1$ \emph{and} $z=\pm 1$. If $\Lambda=M=0$ these solutions 
correspond to the line element:
\[ ds^2=-dt^2+A^2t^2dx^2+dy^2+dz^2 \]
which is easily seen to be locally flat space, if we make the transformation:
\[ \begin{cases} T=t\cosh Ax \\ X=t\sinh Ax \end{cases} \]

The last degenerate case where the Weyl tensor is zero, is only a special 
hypersurface  in space time. It is the hypersurface where
\[ v=\frac{M}{-\Lambda} \]
$(\Lambda<0, M\neq 0)$.

All in all we can say that:
\begin{quote}
\emph{The Weyl curvature invariant, 
$ C^{\alpha \beta \gamma \delta}C_{\alpha \beta \gamma \delta}$, 
will for the family of Bianchi Type $I$ universes with dust and a cosmological
 constant, diverge as $\frac{1}{v^4}$ near the initial singularity, except for
 a set of models of measure zero.} 
\end{quote}This seems to come in conflict with the WCC in its strongest form. 

Wainwright and collaborators~\cite{wa,gw} suggested that it is not the Weyl 
curvature invariant, 
$ C^{\alpha \beta \gamma \delta}C_{\alpha \beta \gamma \delta}$, that would 
represent a ``Gravitational entropy'', but the invariant 
$P^2=\frac{C^{\alpha \beta \gamma \delta}C_{\alpha \beta \gamma \delta}}{
R^{\mu \nu}{R_{\mu \nu}}}$. Let us also check out this more restrictive form 
of the WCC. 

The expression for $P^2$ in a Bianchi type I model is
\[ P^2 =\frac{4}{27}\frac{A^4}{v^2}\frac{1+z^2-2z\cos3\gamma}{M^2+2\Lambda Mv
+4\Lambda^2v^2} \]

If the universe contains dust, this entity diverges as $\frac{1}{v^2}$ near 
the initial singularity. In the absence of dust, this entity diverges as the 
Weyl tensor; the Ricci square  is just a constant. 
Even this more restrictive form of the WCC does not seem to agree with the 
classical Bianchi type I universe.

\subsection{The WCC: Local version}
What conclusions can we draw from these results? Both the entities 
$C^{\alpha \beta \gamma \delta}C_{\alpha \beta \gamma \delta}$ and $P^2$ 
diverge near the origin. Thus the local version of the WCC is set in doubt. 
Based on the above investigations we have to  conclude: 

\begin{enumerate}
\item[$\bullet$] If the WCC is ment to be correct in its strongest form, we 
have to conclude that the classical Bianchi type I model, does not give a good
 description of the initial stages of the universe. The models are too simple 
to take into account the true behaviour of the universe.
\item[$\bullet$] The entity suggested by Wainwright and collaborators are more
 likely to describe a ``gravitational entropy''.

\item[$\bullet$] Including a positive cosmological constant the universe is 
unconditionally driven to isotropization and 
$\frac{ C^{\alpha \beta \gamma \delta}C_{\alpha \beta \gamma \delta}}{ 
R^{\mu \nu}{R_{\mu \nu}}} \longrightarrow 0$ as $v\longrightarrow \infty$.
\end{enumerate}

\section{The Weyl Curvature Conjecture for the Lema\^itre-Tolman models: Local
 version}
Choosing universal time gauge we can write the Lema\^itre-Tolman (L-T) models 
as:
\beq\label{LTmetric}
ds^2=-dt^2+Q^2dr^2+R^2(d\theta^2+\sin^2\theta d\phi^2)
\eeq
where $Q=Q(r,t)$ and $R=R(r,t)$. Classically the metric will be  a solution of
 the Einstein field equations where the spacetime contains a cosmological 
constant $\Lambda$ and dust. The energy-momentum tensor for dust in co-moving 
coordinates is given by $T_{\mu \nu}=\text{diag}(\rho,0,0,0)$. The classical 
equations will now turn into 
\[ R'=FQ\]
where $F=F(r)$ is an arbitrary function, and
\beq\label{meq}
\frac{1}{2}R\dot{R}^2+\frac{1}{2}(1 -F^2)R-\frac{\Lambda}{6}R^3=m
\eeq
The function $m=m(r)$ is given by the integral 
$m(r)=\int^r_0 4\pi\rho R^2 R'dr$, thus can be interpreted as the total mass 
of the dust inside the spherical shell of coordinate radius $r$.
The square root of the Weyl curvature scalar in the L-T models is:
\begin{equation}\label{weylLT}
\sqrt{C^{\alpha \beta \gamma\delta}C_{\alpha \beta \gamma\delta}}=
\frac{2}{\sqrt{3}}\left|\frac{1}{R^2}+\frac{1}{Q^2}\left(\frac{R''}{R}-
\frac{{R'}^2}{R^2}-\frac{Q'R'}{QR}\right)-\frac{\ddot{R}}{R}+
\frac{\ddot{Q}}{Q}-\frac{\dot{Q}\dot{R}}{QR}+\frac{\dot{R}^2}{R^2}\right|
\end{equation}

Using the results obtained in \cite{sigLT} we will investigate the WCC in the 
L-T models. 

\subsection*{The classical value}
Inserting eq. (\ref{meq})  and the classical solution $R'=FQ$ into 
eq. (\ref{weylLT}) we can write the Weyl scalar for all of the L-T models as:
\begin{equation}
\sqrt{C^{\alpha \beta \gamma\delta}C_{\alpha \beta \gamma\delta}}=
{4}{\sqrt{3}}\frac{m}{R^3}\left|1-\frac{m'R}{3R'm}\right|
\end{equation}
If we assume that we can write $R(r,t)=a(t)r$ (which is possible in the FRW 
and the deSitter models), the Weyl tensor will be zero for models where
\[ m\propto r^3 \]
The FRW models do have this behaviour,  thus their Weyl tensor vanishes.  
\par
The Schwarzschild black hole case (which is now in Lema{\^i}tre coordinates) 
has $m'=0$ and the Weyl tensor squared consists solely of the Kretschmann 
scalar. The square root of the Weyl scalar diverges as $R^{-3}$. 

Let us define a mean dust density function $\bar{\rho}$ by the relation:
\begin{equation}
m(r)=\frac{4}{3}\pi\bar{\rho}R^3
\end{equation}
Then we can write the Weyl curvature invariant as:
\[ 
C^{\alpha \beta \gamma\delta}C_{\alpha \beta \gamma\delta}=
\frac{16^2}{3}\pi^2(\bar{\rho}-\rho)^2 \]
$m(r)$ is a function of $r$ only, so
\[ \frac{\dot{\bar{\rho}}}{\bar{\rho}}=-3\frac{\dot{R}}{R}\]
The physical meaning of the Weyl curvature invariant is now clear; it measures
 the difference between the actual dust density and the mean dust density. If 
$\rho=\bar{\rho}$ everywhere, we have a FRW universe with a cosmological 
constant, thus a zero Weyl tensor.  
Iff the Weyl curvature invariant should be ever increasing then the function 
$d\equiv \bar{\rho}-\rho$ has to obey: $d\dot{d}>0$.

Let us check out the Weyl tensor more explicitly. Near the initial singularity
 (${R\longrightarrow 0}$) we can write the solution of the L-T model for all 
the models approximately as:
\begin{equation} \label{smallimit}
R\approx \left(\frac{9}{{2}}m\right)^{\frac{1}{3}}(t-t_0(r))^{\frac{2}{3}}
\end{equation}
which leads to 
\beq \label{rhodiff}
4\pi(\bar{\rho}-\rho)=
\frac{3m}{R^3}\cdot\frac{2t'_0}{2t'_0-\frac{m'}{m}(t-t_0)} 
\eeq
To avoid intersecting world-lines we have to assume $t'_0(r)<0$ \cite{miller} 
which makes the difference $\bar{\rho}-\rho$ positive. 
 The above entity will therefore diverge as $R \longrightarrow 0$ unless 
$t'_0=0$ which is the FRW case. \emph{ Thus, the Weyl scalar invariant will 
diverge when  $R\longrightarrow 0$ in the generic L-T model.}
\par
Let us also check out the scalar suggested by Wainwright and collaborators.
The Ricci square in the classical L-T model is:
\[R^{\mu \nu}R_{\mu \nu}= 4\Lambda^2+16\pi\rho\Lambda+64\pi^2\rho^2\]
We can therefore write
\begin{equation}
P^2 =\frac{4}{3}\frac{\left(\frac{\bar{\rho}}{\rho}-1\right)^2}{\left(
\frac{\Lambda}{4\pi\rho}\right)^2+\left(\frac{\Lambda}{4\pi\rho}\right)+1}
\end{equation}

In the small $R$ limit we can use the preceding solution. The $\Lambda$ term 
may be neglected ($\rho$ diverge), so near the initial singularity the entity 
behaves as
\begin{equation}
P^2 \propto \left(\frac{2mt'_0}{m'(t-t_0)}\right)^2
\end{equation}
 Thus, we have to conclude that (unless $t'_0=0$)
\begin{equation}\label{wainbeh}
\frac{\partial P^2}{\partial t}\bigg{|}_{R\longrightarrow 0}<0
\end{equation}
Bonnor~\cite{Bon85} considered the model $1-F^2>0$ with a zero cosmological 
constant upon which he studied the behaviour of the same entity. He did 
however reach the opposite result. But as he clearly points out, he inserted 
the condition $t'_0=0$ \emph{by hand}. He assumed an isotropic initial 
singularity, thus demanding that the Weyl tensor is zero initially. As we see,
 with a more general assumption the entity suggested by Wainwright and 
collaborators will initially be a \emph{decreasing} function of time. 

To see what is actually happening if we push $t'_0(r)$ towards zero in the 
case $1-F^2>0$, we define a new time variable $T=t-t_0(r)$. Now a dot means 
derivative with respect to $T$. Close to $T=0$ we can write 
\[ 
R^3=\frac{9}{2}mT^2g(r,T)
\]
where $g(r,T)\approx 1-\epsilon(r)T^{\frac{2}{3}}$ and $\epsilon(r)>0$. 
The entity $P^2$ now becomes
\beq\label{compb}
P^2 \propto \frac{1}{(gm')^2}\left[g'-t'_0\left(2\frac{mg}{T}+
m\dot{g}\right)\right]^2
\eeq

For $t_0(r)$ small but non-zero $P^2$ will diverge at the origin. As $T$ 
increases $P^2$ will rapidly decrease, until the $g'$ part becomes dominant in
 eq. (\ref{compb}). $P^2$ reaches a minimum before it will increase again. As 
$t'_0$ is driven towards zero, this minimum approaches zero, and at $t'_0=0$, 
the minimum will be at $T=0$ where $P=0$. We see that the behaviour at the 
initial singularity is drasticly altered near $t'_0 =0$.

\subsection*{Large $R$ limit}
In the large $R$ limit and in the presence of a cosmological constant, we have
 the deSitter-like solutions. As $R$ heads for infinity, we can approximate 
these solutions as $R\propto e^{Ht}$. The ratio $\frac{\bar{\rho}}{\rho}$ is 
then approximately constant. The Wainwright entity will therefore go as
\[P^2 \propto R^{-6}\approx e^{-6Ht} \]
It is exponentially decreasing as a function of $t$, and at the end of the 
inflationary era, it is close to zero. 

In a $1-F^2=0$ matter dominated universe, we can set $\Lambda\approx 0$ to 
obtain the solutions of eq. (\ref{smallimit}). The behavior of the entity 
$P^2$ is now exactly the same as for the small $R$ limit, but the relation 
(\ref{wainbeh}) will hold for any $R$. Even in this case the WCC is set in 
doubt.

\section{Gravitational Entropy: Revisited}

As we have seen the entity $P^2$ diverges badly at the initial singularity, 
and decreases as $(t-t_0)^{-2}$ shortly after the initial singularity in the 
L-T model. Firstly this seems rather contra-intuitive. The Weyl tensor squared
 measures the square of the difference between the mean dust density 
$\bar{\rho}$ and the actual dust density $\rho$. However, both of these 
densities diverge at the initial singularity. Shortly after the Big Bang both 
are reduced as the universe expands, and $\rho$ is only reducing as 
$(t-t_0)^{-1}$ compared to $\bar{\rho}\propto (t-t_0)^{-2}$. If the 
gravitational entropy is infinite at the Big Bang singularity, the total 
entropy also has to be infinite, which is contrary to the second law of 
thermodynamics. 

Let us consider a finite 3-volume $V$ in our space time. According to the 
first law of thermodynamics the matter entropy $S_M$, the internal energy $U$ 
and the number of particles $N$ in $V$ will evolve as:
\beq
TdS_M=dU+pdV-\mu dN
\eeq
If this volume is co-moving then $dN=0$. If the matter content is dust then 
$p=0$. 
There are a couple of things to note. Firstly in a dense dust cloud, we expect
 the internal energy of the dust to be high, thus we expect the entropy to be 
high. Secondly the entropy is an increasing function of the volume. 
Let us therefore consider a co-moving volume $V$ in our spacetime. What could 
the expression for a {\it gravitational entropy} be? From the previous 
sections we noticed that an L-T model with $\bar{\rho}=\rho$ is homogeneous. 
The quantity $\frac{\bar{\rho}-\rho}{\rho}$ is an inhomogeneity measure in the
 L-T models. In the absence of a cosmological constant we notice that 
\beq
P=\frac{2}{\sqrt{3}}\left(\frac{\bar{\rho}-\rho}{\rho}\right)
\eeq
The sign of $P$ is here chosen so that the  configuration 
$\bar{\rho}<\rho$ is associated with a $P<0$ while the more realistic 
configuration $\bar{\rho}>\rho$ has $P>0$ corresponding to \ref{rhodiff} 
positive. The matter entropy is a function of
 the volume $V$, thus the gravitational entropy ought to be so too. We 
therefore consider the entity defined by:
\beq 
{\mathcal S}=\int_V P dV
\eeq
Introducing co-moving coordinates $x^i$ we write $dV=\sqrt{h}d^3x$ where 
$\sqrt{h}$ is the 3-volume element. The integration range is now constant as a
 function of time and if we integrate over a unit coordinate volume which is 
so small that the integrand is approximately constant, we may write 
\[ {\mathcal S}=\int_V P dV\approx P\sqrt{h}\]

This volume can in principle be as small as possible and in this sence we can 
treat the entity $\Pi\equiv\sqrt{h}P$ as a local entity. This point was 
emphasized by Hayward\cite{hayward}. Following Hayward we can describe entropy 
by an entropy current vector $\vec{\Psi}$ with the second law being
\beq
\nabla\cdot\vec{\Psi}\geq 0
\eeq
where  $\vec{\Psi}$ is future-casual. Decomposing  $\vec{\Psi}$ into 
components tangent and orthogonal to the material flow vector $\vec{u}$ 
($\vec{u}\cdot \vec{u}=-1$):
\beq 
\vec{\Psi}=s\vec{u}+\vec{\varphi}
\eeq
where $\vec{u}\cdot\vec{\varphi}=0$, $s$ is the entropy density and 
$\vec{\varphi}$ is the entropy flux. 

In our case we take the gravitational entropy density proportional to $P$ and 
assume a vanishing entropy flux:
\beq
\vec{\Psi}\propto P\vec{u}
\eeq
 and identify $\vec{u}$ with the cosmological time vector. Thus 
\beq
\vec{u}\cdot\nabla P+P\nabla\cdot\vec{u}\geq 0.
\eeq
By the identity $\vec{u}\cdot \nabla\sqrt{h}=\sqrt{h}\nabla\cdot \vec{u}$ we 
get
\beq\label{law2}
\vec{u}\cdot\nabla(\sqrt{h}P)=\vec{u}\cdot\nabla \Pi\geq 0
\eeq

Thus in this sence we can treat $\Pi$ as a local entity. The equation 
\ref{law2} says that the directional derivative of $\Pi$ along the material 
flow vector are everywhere positive. In our case this simply reduces to
\[ \frac{\partial}{\partial t}\Pi \geq 0 \]
If we want to relate $\Pi$ to the entropy density as above we will expect an 
increasing $\Pi$ as a function of time. 
The difference in ``localness'' of $P$ and $\Pi$ is that $P$ is a scalar or a 
{\it 0-form}, while $\Pi$ transforms as the components of a differential form 
of maximal rank i.e. of a form that can be written 
$\omega=\Pi \bigwedge_i dx^i$ where the index runs over the dimension of the 
manifold\footnote{Or simply $\omega=\pm\star P$ where $\star$ is the Hodge 
star operator.}. We will however in this article use the ``integrated'' 
version of $\Pi$ and therefore continue to call the entity $\Pi$ non-local 
even though it the above sense might as well be called local.

In both articles \cite{sigBI,sigLT} we considered co-moving dust so we can use
 these results directly. We will therefore investigate the entity 
\beq 
\Pi=P\sqrt{h}=\pm \sqrt{h}\left(\frac{C_{\alpha \beta\gamma\delta}C^{\alpha 
\beta\gamma\delta}}{R_{\mu\nu}R^{\mu\nu}}\right)^{\frac{1}{2}}
\eeq
In the L-T model we can choose the sign $\pm$ as mentioned above. We will 
however consider only the case where $\Pi\geq 0$.

\subsection{Behaviour of $\Pi$ in the L-T model}
In the L-T model the entity  $\Pi$ turns into
\beq 
 \Pi=\frac{P}{F}R^2R'
\eeq
In the small $R$ limit we can ignore the $\Lambda$ term. If $\Lambda=0$, $\Pi$
 is simply
\[ \Pi=\frac{2R^3}{\sqrt{3}F}\frac{|\bar{\rho}'|}{\rho} \]

In the limit $R\longrightarrow 0$ we get:
\beq
\Pi=\frac{8}{3\sqrt{3}}\frac{m|t_0'|}{Fm'}(m'(t-t_0)-2t'_0)
\eeq
which is finite for $t-t_0=0$ and increasing thereafter. This entity has the 
right behaviour. 

In the deSitter limit with $\Lambda>0$ the universe is expected to approach 
homogeneity. In this limit the $\Pi$ will go as
\beq
\Pi=\frac{2}{\sqrt{3}}\frac{m'}{F\Lambda}\left(\frac{\bar{\rho}-
\rho}{\rho}\right)
\eeq
Since $\bar{\rho}$ will approximately have the same time evolution as $\rho$ 
in the deSitter limit, $\Pi$ will asymptotically evolve towards a constant. 
The solutions in the deSitter limit can be written as 
$R=\exp\left(\sqrt{\frac{\Lambda}{3}}t+{\frac{m}{3}f(r)}\right)$ where $f(r)$ 
is some function of $r$ determined by eq. (\ref{meq}). The expression for 
$\Pi$ is now 
\[\Pi= \frac{2}{\sqrt{3}}\frac{{m'}^2m^2}{F\Lambda}f'(r) \]
The physically realistic cosmologies has $f'(r)>0$ thus $f(r)$ is an 
increasing function of $r$. The relation between  $f(r)$, $m$, $F(r)$ and 
$t_0(r)$ is not calculated here but can be found by using the result obtained 
by Zecca\cite{Zecca}. The qualitative behaviour is, however, expressed in the 
present relation.

It appears as if the deSitter state is a stable state. Recall that for a 
stable thermo-dynamical state, the entropy will be constant. Interestingly 
since $\Pi\neq 0$ in the deSitter limit, some of the information of the 
initial state is present. 

The evolution of the entity $\Pi$ for the L-T model can be summarized:
\begin{enumerate}
\item[$\bullet$]{\bf Large $\Lambda$ and ever expanding:} In the initial epoch
 the dust dominates and  $\Pi$ is increasing linearly in $t$. The universe is 
becoming more and more inhomogeneous. After the universe has grown  
considerably, the cosmological constant becomes dominant, $\Pi$ stops growing 
and if the cosmological constant is large enough, it decreases asymptotically 
towards a constant value. The universe is smoothened out. 
\item[$\bullet$]{\bf Small $\Lambda$ and ever expanding:} Again the dust 
dominates initially. The cosmological constant is too small to make $\Pi$ 
decreasing. The entity  $\Pi$ is ever increasing but is bounded from above by 
a relatively large constant value.
\item[$\bullet$]{\bf Zero $\Lambda$ and ever expanding:} The  $\Pi$ will again
 be ever increasing and will asymptotically move towards a function 
$f(t)=c+bt^p$ where $c$ and $b$ are constants and $p=3$ iff $F^2>1$ and $p=1$ 
iff $F^2=1$.
\item[$\bullet$]{\bf Recollapsing universe:} Due to the dust term, the final 
singularity will not be symmetric to the initial singularity.   Hence this 
entity is asymmetric in time for a recollapsing universe.
\end{enumerate}

In the L-T models we see that $\Pi$ behave just like the WCC suggests. 

The Schwarzschild spacetime has $m(r)=constant$ and has a vanishing Ricci 
tensor. If we look at the entity $\Pi$ in the region outside the Schwarzschild 
singularity $\Pi$ will diverge. This is in some sence the maximal possible 
value of $\Pi$, the Weyl tensor is as large as possible and the Ricci tensor 
is the smallest as possible. Thus at this classical level it seems that the 
Schwarzchild spacetime has the largest possible $\Pi$ which is a good thing if 
one wants to connect $\Pi$ with the entropy of the gravitiational field. 
\subsection{$\Pi$ in the Bianchi type I model}
We can now use the same entity $\Pi$ for the Bianchi type I model which we 
``derived'' for the L-T model. The expression for $\Pi$
in the Bianchi type I model is
\beq
\Pi=\sqrt{h}P=\frac{2A^2}{3\sqrt{3}}\left(\frac{1+z^2-2z\cos3\gamma}{M^2+
2\Lambda Mv+4\Lambda^2v^2}\right)^{\frac{1}{2}}
\eeq
where $\sqrt{h}=v$.
Near the initial singularity we can make a Taylor expansion of $\Pi^2$ to 
first order in $v$:
\[
\Pi^2\approx \frac{8}{27}\frac{A^4}{M^2}(1-\cos3\gamma)\left(1+
\left(\frac{3M}{2A^2}-\frac{2\Lambda}{M}\right)v\right)
\]
Interpreting $m=\frac{M}{A}$ as the dust density\cite{sigBI} in coordinate 
space, we see that as $v\ra 0$, $\Pi\ra \frac{2A}{3\sqrt{3}m}$. But $\Pi$ only
 increases iff $3m^2>4\Lambda $. Thus if $m\neq 0$ this inequality is 
satisfied for $\Lambda\leq 0$. Investigating the classical solutions for 
$\Lambda>0$ more closely we see that iff $3m^2=4\Lambda$, the volume element 
$v(t)$ can be written as a part of the exponential map. In the case where 
$3m^2>4\Lambda$, $v(t)$ can be written as a part of the hyperbolic cosine and 
if $3m^2<4\Lambda$, $v(t)$ can be written as a part of the hyperbolic sine. 
Thus if the solutions expand too rapidly, 
the dust term does not manage to increase the value of $\Pi$ immediately 
after the initial singularity.

Also in the Bianchi type I case the deSitter limit will yield a constant value
 of $\Pi$:
\[
\lim_{v\ra\infty}\Pi=\frac{A}{3\Lambda^{\frac{1}{2}}}
\]

In the absence of a cosmological constant the result is simply:
\[
\Pi=\frac{2A}{3\sqrt{3}m}\left((2+3mt)(1-\cos 3\gamma)+
\frac{9}{4}m^2t^2\right)^{\frac{1}{2}}
\]
In this case $\Pi$ is monotonically increasing. For large $t$ we have 
$\Pi\approx \frac{A}{\sqrt{3}}t$, which is according to the same power-law as 
in the ``flat'' and $\Lambda=0$ case of the L-T model.

To summarize our investigation of $\Pi$ in the Bianchi type I model we can say
 that the entity $\Pi$ at the initial singularity is a constant determined by 
the inverse of the dust density. For $3m^2>4\Lambda$ it will increase 
immediately after the initial singularity. In most cosmological considerations
 it is assumed that the cosmological constant is small. The exception is in 
the inflationary era in which the vacuum energy dominates over  all other 
matter degrees of freedom. The inflationary era will smooth out anisotropies 
as well as inhomogeneities, and the behavior of $\Pi$ in this case is 
therefore expected. In the deSitter limit $\Pi$ will asymptotically move 
towards a constant. Large $\Lambda$ means small value, while small $\Lambda$ 
corresponds to a large value. This is in full agreement with the L-T models. 
It is also interesting that the entity $\Pi$ is very sensitive to different 
matter configurations. This makes it a lot easier to check whether the $\Pi$ 
has the right entropic behaviour. 

It also appears to us that the Kasner universe is a sort of ``Worst case 
scenario''. It has a horrendous initial singularity, a cigar-shaped 
singularity where the solution space is completely disconnected from the 
isotropic FRW universe. Also the apparently flat case, $\cos 3\gamma =1$, 
possess a severe and devastating non-smooth singularity at $t=0$. In this case
 the singularity is that of a cone and is a true singularity in the smooth 
structure of the space time\cite{sigBI}. The entity $\Pi$ is for $v=0$ a 
continuous function in $\gamma$ so the $\cos 3\gamma =1$ case is not 
particularly different from the cases $\cos 3\gamma \neq 1$. It seems that 
while the Schwarzschild black hole case serves as an upper bound in the 
inhomogeneous case the Kasner universe serves as an upper bound for the 
anisotropic cosmologies. 

\begin{figure}
\centering
\subfigure[$\cos 3\gamma=-1$]{\epsfig{figure=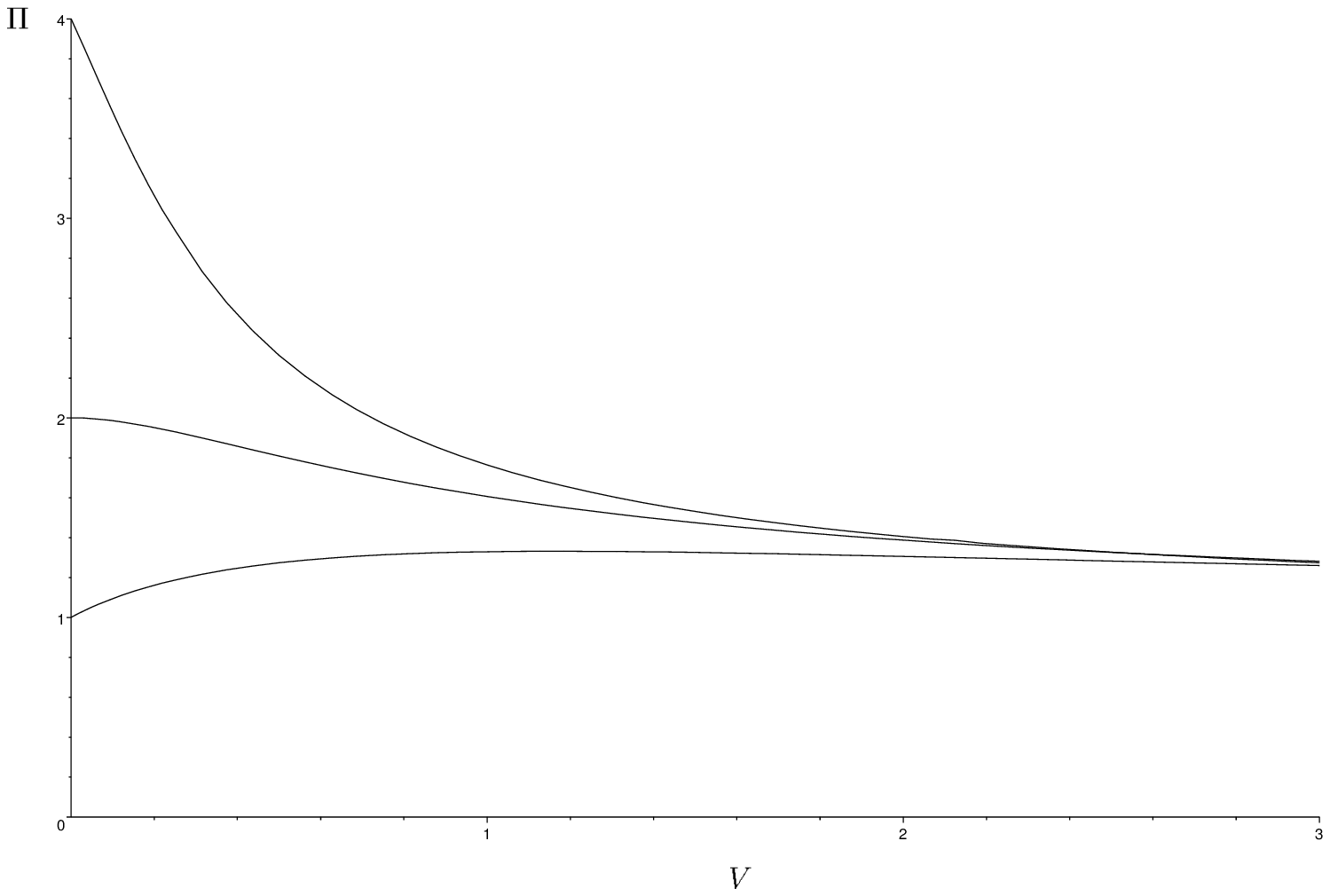, width=8cm}}
\subfigure[$\cos 3\gamma=\frac{1}{2}$]{\epsfig{figure=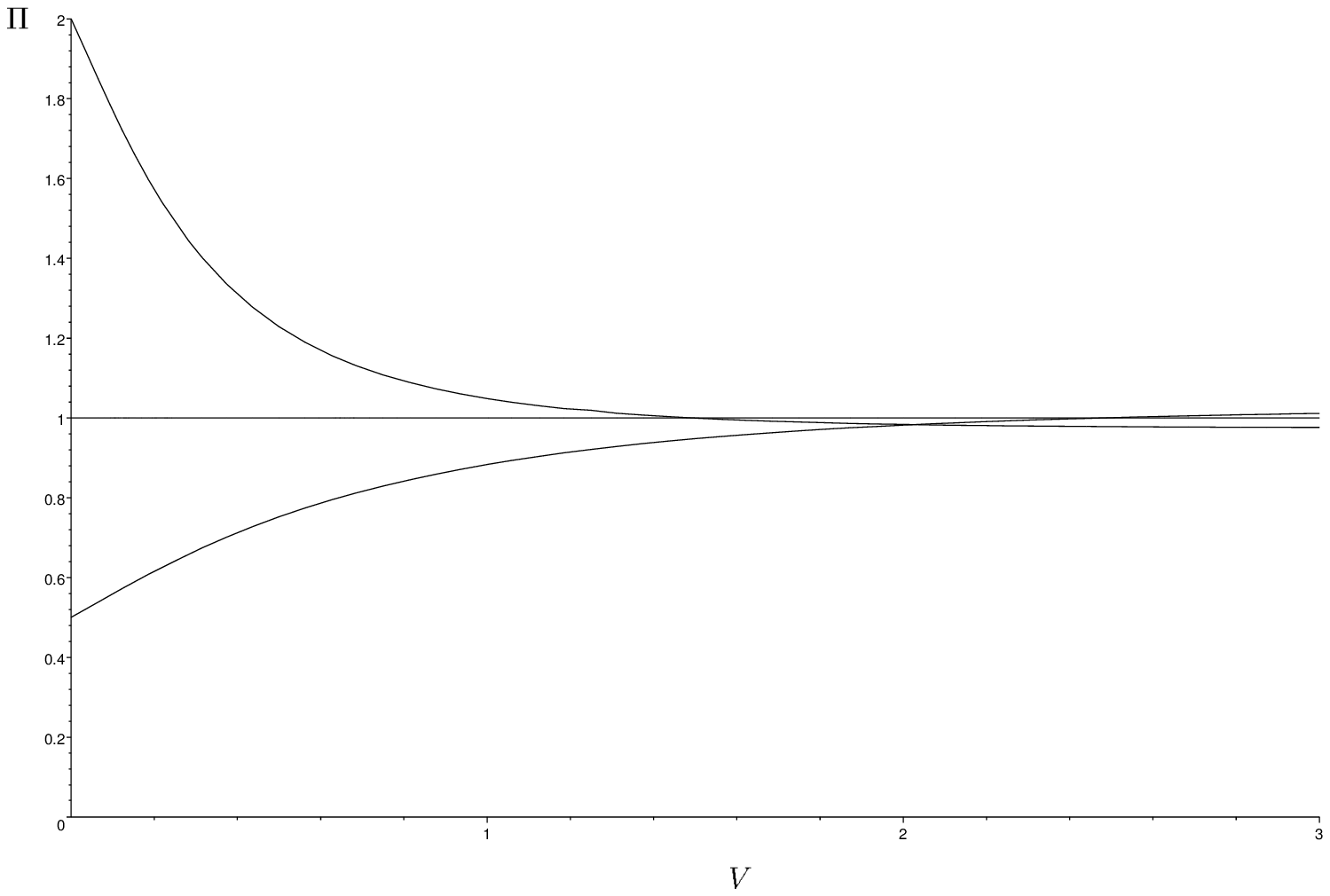, width=8cm}}
\subfigure[$\cos 3\gamma=1$]{\epsfig{figure=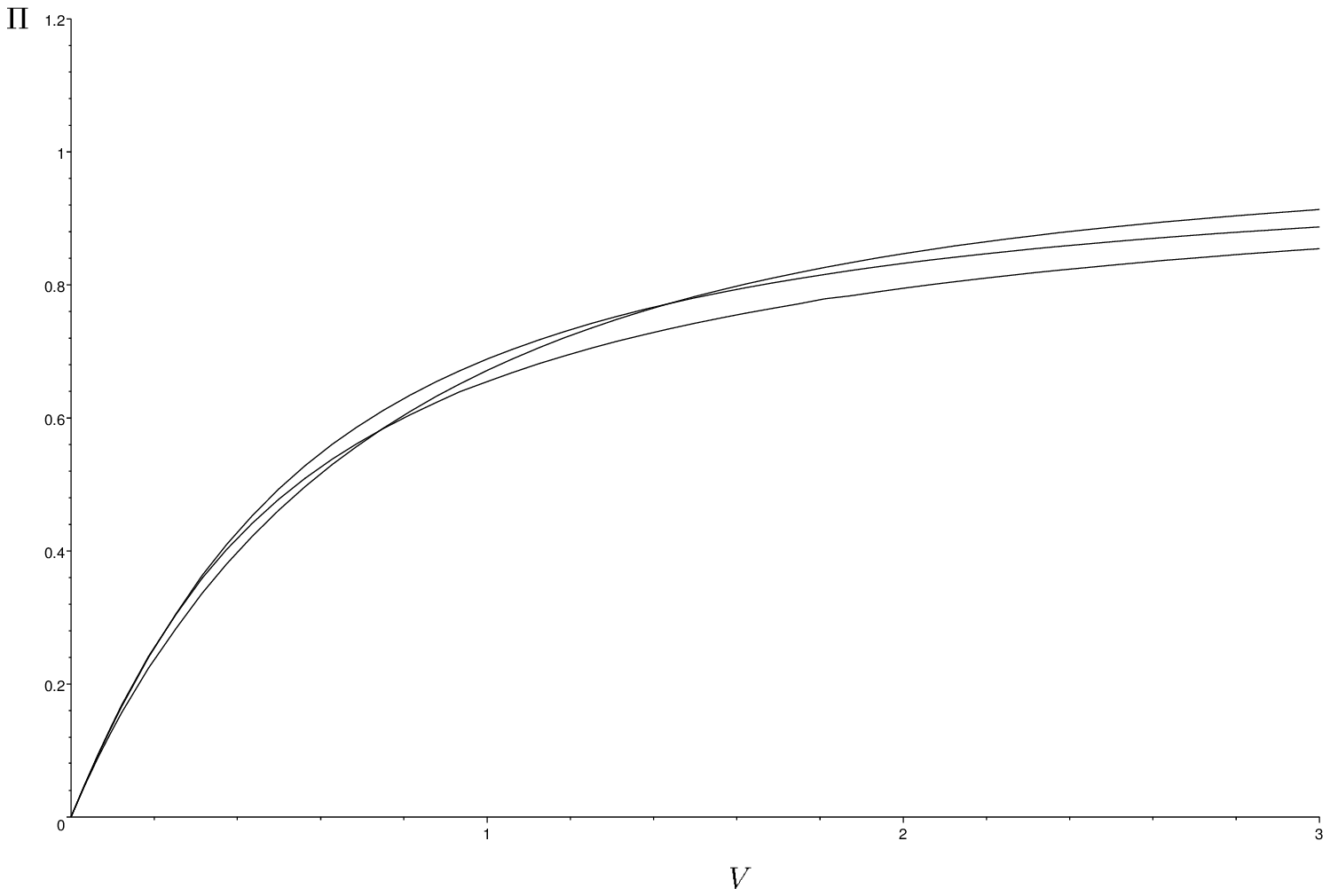, width=8cm}}
\caption{Plots of $\Pi(V)$ for the Bianchi type I model with different values 
of $\cos 3\gamma$. The cosmological constant is set to $\Lambda=\frac{3}{4}$ 
and in each plot graphs with $m=\frac{1}{2}, 1$ and 2 are drawn. We have also 
chosen $A=\frac{3}{2}\sqrt{3}$ so that for all of the above cases $\Pi\lra 1$ 
in the limit $V\lra \infty$.}\label{BianchiIpos}
\end{figure}

In figure \ref{BianchiIpos} we have plotted $\Pi$ as a function of 
$V\equiv \frac{v}{A}$ in the case of a positive cosmological constant. By 
investigating the entity $\Pi$ we note that there is one particular case 
where $\Pi$ is a constant as a function of $V$. It is the case where 
$3m^2=4\Lambda$ and $\cos 3\gamma=\frac{1}{2}$ and it corresponds to a line 
element that can be written\cite{sigBI}:
\begin{align}\begin{split}\label{eq:genr}
ds^2=&-\frac{d\tau^2 }{\left(1-\sqrt{3\Lambda}\tau\right)^2} \\ 
&+\frac{1}{\left(1-\sqrt{3\Lambda}\tau\right)^{\frac{2}{3}}}
\tau^{\frac{2}{3}}\left[\tau^{\frac{4}{3}\cos(\frac{\pi}{9})}dx^2+
\tau^{\frac{4}{3}\cos(\frac{5\pi}{9})}dy^2+
\tau^{\frac{4}{3}\cos(\frac{7\pi}{9})}dz^2 \right]
\end{split}\end{align}
 . 

\begin{figure}
\centering
\subfigure[$\cos 3\gamma=-1$]{\epsfig{figure=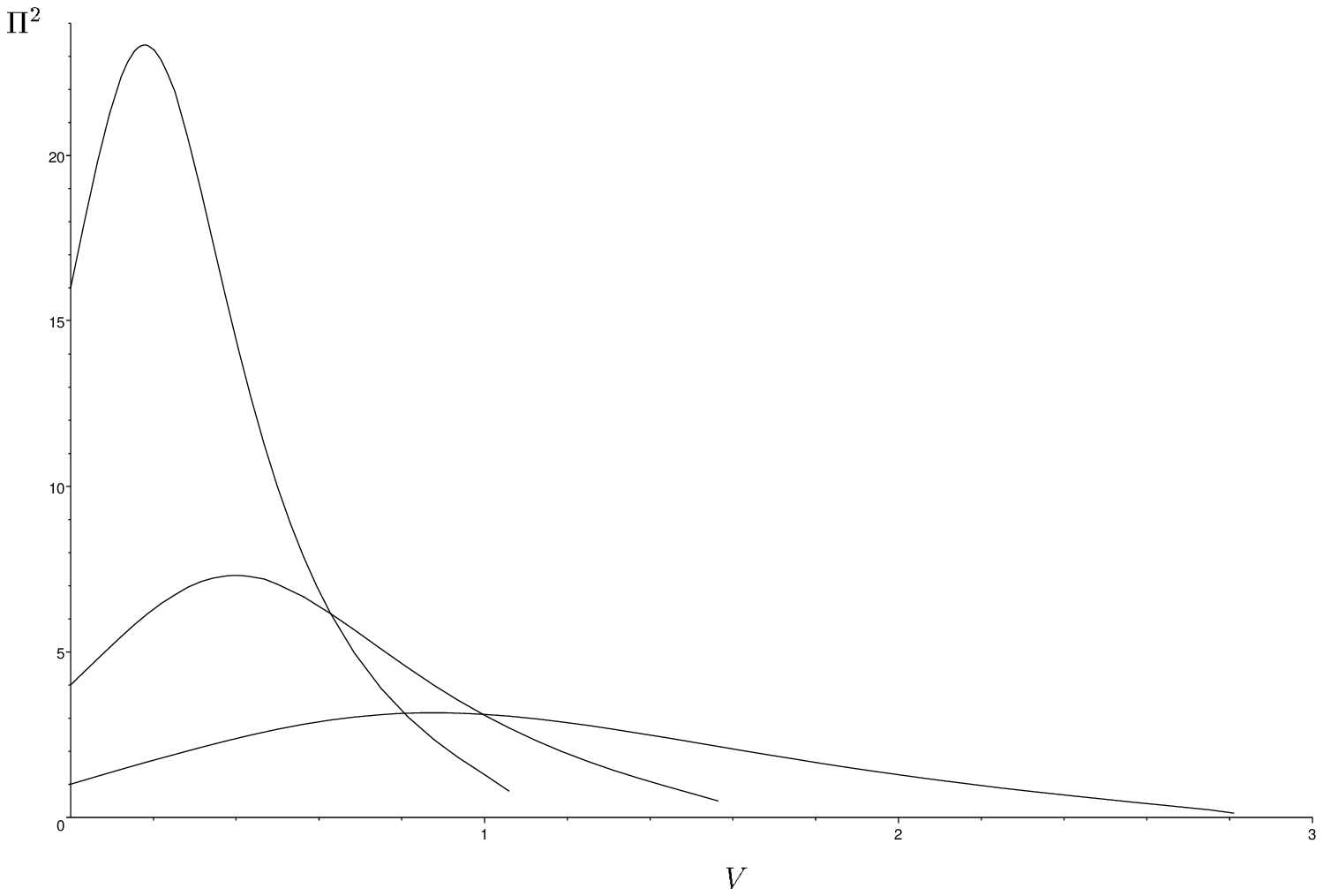, width=8cm}}
\subfigure[$\cos 3\gamma=1$]{\epsfig{figure=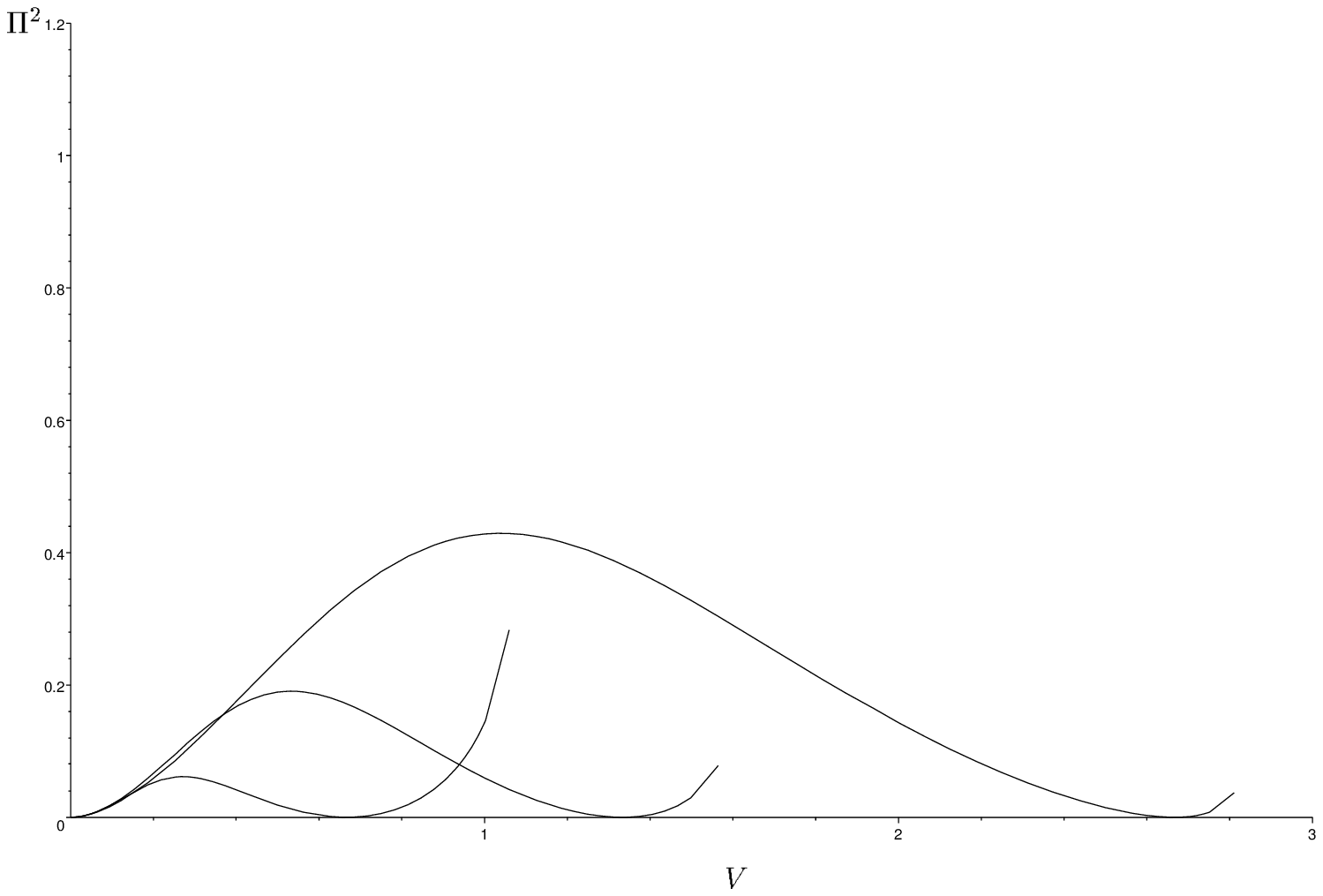, width=8cm}}
\subfigure[$\cos 3\gamma=-\frac{1}{2}$]{\epsfig{figure=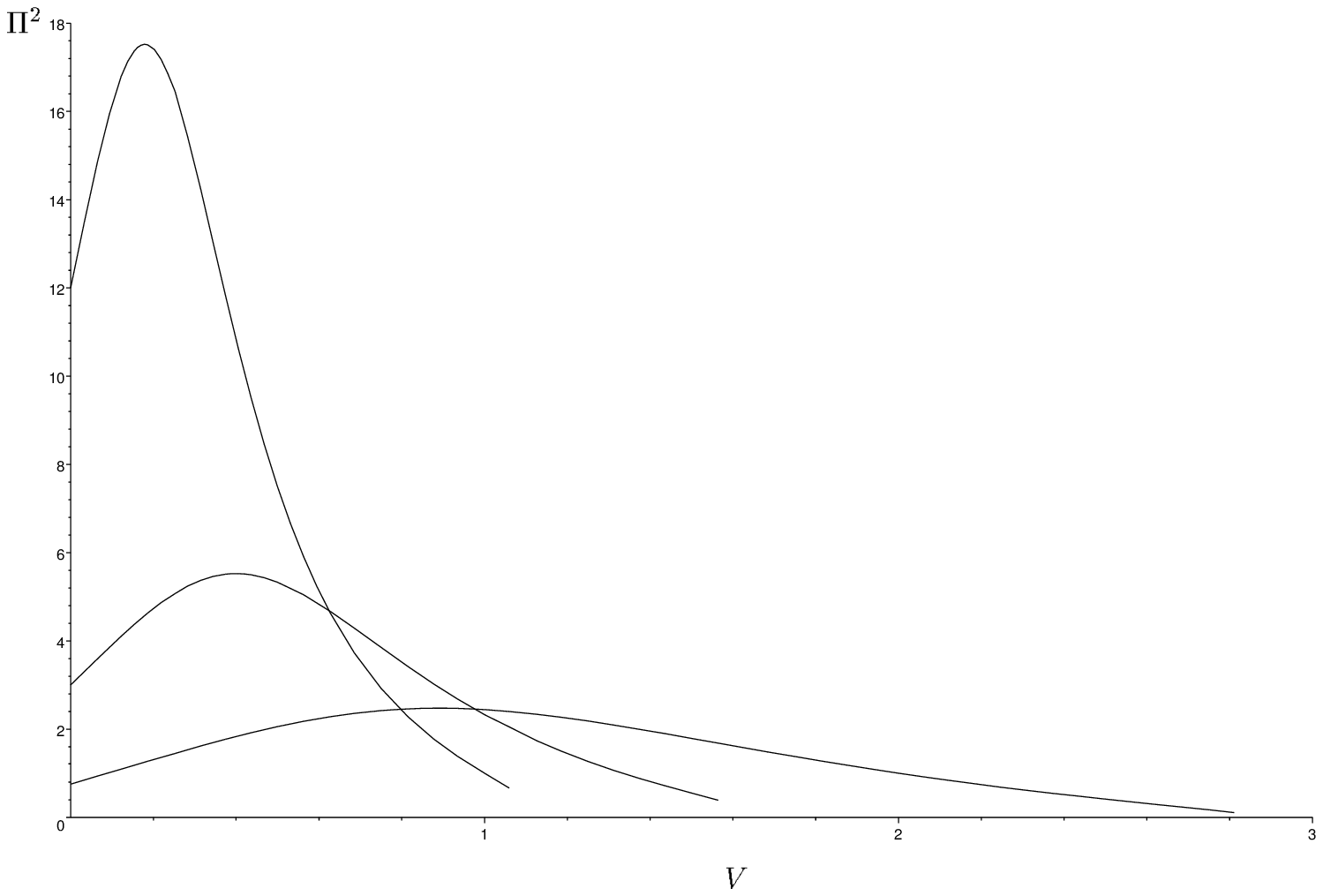, 
width=8cm}}
\subfigure[$\cos 3\gamma=\frac{1}{2}$]{\epsfig{figure=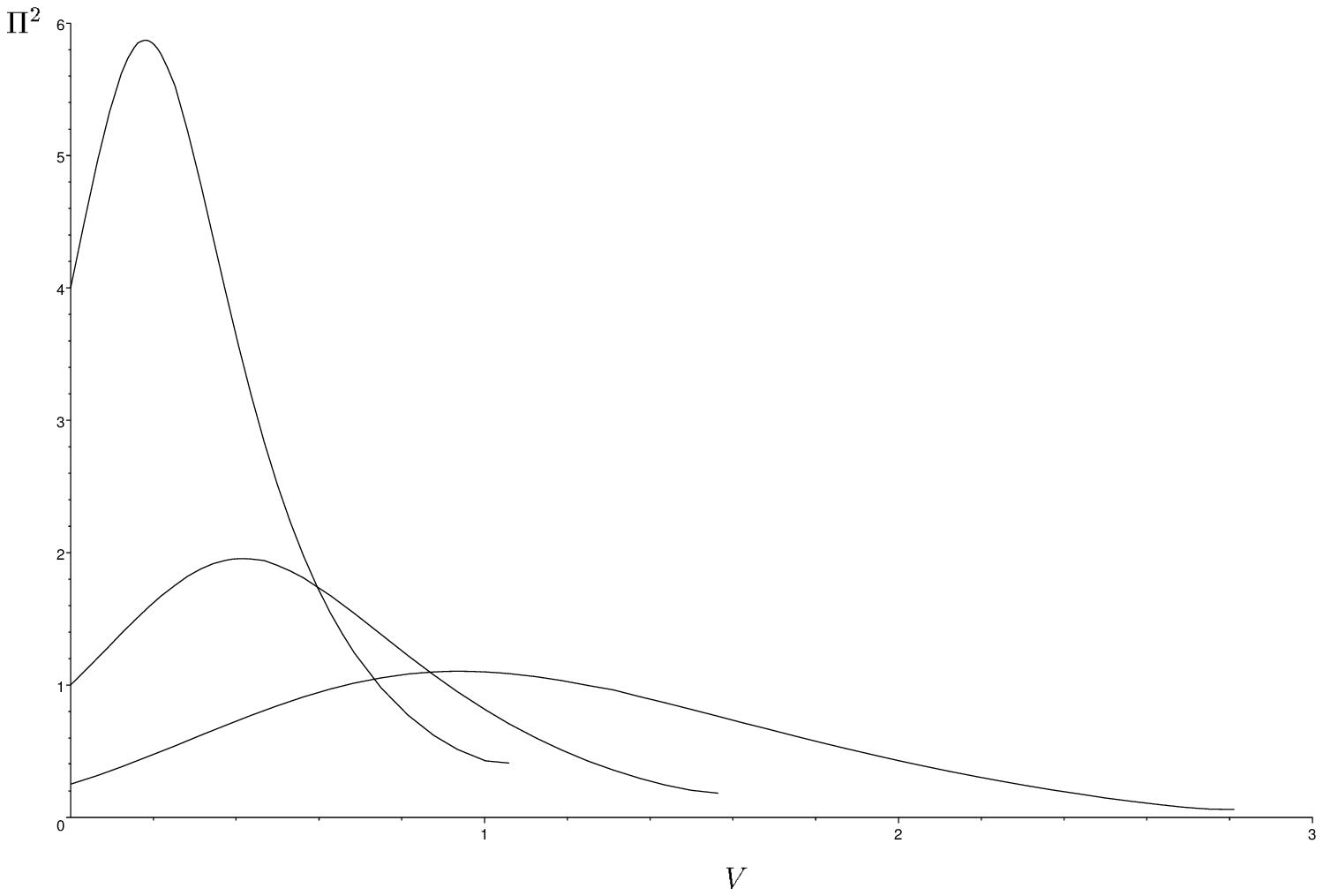, width=8cm}}
\caption{Plots of $\Pi(V)^2$ for the Bianchi type I model with different 
values of $\cos 3\gamma$. The cosmological constant is set to 
$\Lambda=-\frac{3}{4}$ and in each plot graphs with $m=\frac{1}{2}, 1$ and 2 
are drawn. The expanding phase can be mapped onto the contracting phase by 
changing the sign of $\cos 3\gamma$. $A$ is chosen to be equal to 
$\frac{3}{2}\sqrt{3}$.}\label{BianchiIneg}
\end{figure}

In figure \ref{BianchiIneg} we have plotted $\Pi^2$ as a function of $V$ in 
the case of a negative cosmological constant. The expanding solutions can be 
mapped onto the contracting solutions by changing the sign of $\cos 3\gamma$. 
Thus if $\cos 3\gamma=0$ the entity $\Pi$ will be symmetric with respect to 
the turning point where the universe turns from an expanding to a contracting 
phase. We note that only if $\cos 3\gamma >0$ the function $\Pi$ will be 
larger in the contracting phase than in the expanding phase. The case where 
$\cos 3\gamma=0$ is symmetric with respect to the turning point of the 
evolution of the Bianchi type I universe. Thus one might say that this case 
is the most symmetric of the Bianchi type I universes, while the cases 
$\cos 3\gamma =\pm 1$ are the most asymmetric cases conserning the time 
evolution\footnote{If we look at the spacetimes these solutions correspond to,
 we note something interesing. The case $\cos 3\gamma=1$ Kasner universe 
corresponds to the conar-like universe described in \cite{sigBI}. The (local) 
symmetry group of the spatial section of all these related spacetimes (with 
matter content) can be expanded to 
$\mb{R}\times\text{Sym}(E^2)=\mb{R}\times ISO(2)$. This is also the case for 
the solutions described by $\cos 3\gamma=-1$. These solutions lie just on the 
opposite side of the Kasner circle from the conar-like universes 
$\cos 3\gamma =1$. Thus these solutions (the conar-like universe 
$\cos 3\gamma=1$, and the ``anticonar''-like universe $\cos 3\gamma =-1$) 
have actually a larger (local) symmetry group than the other solutions 
$\cos 3\gamma \neq \pm 1$.}.    
\section{Quantum cosmology and the Weyl curvature conjecture}
In the previous section we investigated the entity $\Pi$ in the classical 
cosmologies of the L-T model and in the Bianchi type I model. We saw that is 
had a very promising behaviour. What can the theory of Quantum Cosmology say 
about its behaviour? Is a small initial value of $\Pi$ more likely than a 
large value? Again we will use the results from the two papers 
\cite{sigBI,sigLT} and we will start by investigating the L-T models.

\subsection{The L-T model}

The full wave function for the universe is a linear combination of particular 
solutions of the Wheeler-DeWitt equation:
\[ \Psi=\sum_iC(i) \rho(i)\Psi_i \]
where $i$ runs over some index set. In the paper \cite{sigLT} we considered 
semi-classical tunneling wave functions. The universe was tunneling from a 
matter-dominated universe which was classically confined to a finite size, 
into a deSitter like universe. After the tunneling across the classically 
forbidden region the universe described vacuum-dominated models, similar to an
 inflationary model of the universe. Thus in the semi-classical approximation 
the wave function will have a superspace current which  will flow 
asymptotically towards deSitter-like solutions. 

The actual expectation values from entities like 
$C^{\alpha\beta\gamma\delta}C_{\alpha\beta\gamma\delta}$, $P^2$ and $\Pi$ are 
not explicitly obtained for these models, because the actual calculations 
suffer from ``endless'' expressions and highly time-consuming quantities. We 
will therefore give a more general description of the evolution of the Weyl 
tensor for the L-T models. 

It would be useful to first recapitulate some of the discussion done in 
\cite{sigLT}. First of all we discussed tunneling wavefunctions in the WKB 
approximation. In the WKB approximation we assume that the wave function has 
the form 
$\Psi_{WKB}=\exp(\pm iS)$,
to the lowest order we got the Hamilton-Jacobi equation: 
\begin{equation}\label{energidiff}
\left(\frac{\delta S}{\delta R}\right)^2-
\frac{{F'}^2}{F^4}\left[2mR-R^2(1-F^2)+\frac{\Lambda}{3}R^4\right]=0
\end{equation}
If we assume that $S=\int \sigma(r) dr$ the resulting equation will be  
the Hamilton-Jacobi equation for a point particle with action $\sigma(r)$
($r$ is only a parameter). In the Hamilton-Jacobi equation the functional 
$S$ turn out to be the action at the classical level. Since the classical
 action can be written as an integral over $r$ the assumption 
$S=\int \sigma(r) dr$ is therefore reasonable at the lowest order WKB 
level. 
We can interpret the action $\sigma$  as the action of a point particle 
moving in a potential 
$V(R)=\frac{{F'}^2}{F^4}\left[-mR
+\frac{1}{2}(1-F^2)R^2-\frac{\Lambda}{6}R^4\right]$ with zero energy. The 
WKB wavefunction $\psi_{WKB}$ for the point particle can then be written
$\psi_{WKB}=\exp(\pm i\sigma)$. The two WKB wavefunctions can therefore 
be 
related by $\Psi_{WKB}=\exp(\int dr\ln \psi_{WKB})$. Finding first the 
wavefunction $\psi$ we can then relate its WKB approximation with 
$\Psi_{WKB}$ through  $\Psi_{WKB}=\exp(\int dr\ln \psi_{WKB})$.

Let us now ask the question: \emph{Is it more likely for a universe with small 
Weyl tensor to tunnel through the classical barrier than a universe with a 
large Weyl tensor?} If this was the case this would have been in agreement 
with the WCC. The question is difficult to answer in general but we shall make
 some simple considerations in order to shed some light upon it. 

We assume that the dust density near the origin of the coordinates is larger 
than further out. The homogeneous mass function goes as 
$m_h(r)=\frac{4}{3}\pi\rho_h r^3$ where $\rho_h$ is a constant. If the 
dust density is larger near the origin of the $r$-coordinate than for larger 
values of $r$ then $m(r)\geq m_h(r)$ if we demand equality only 
at $r=r_{max}$(we only look at closed universes). This will not in general 
change the size of the universe so we can look at the effects from $m(r)$ 
alone. 
Since $m(r)$ is greater in general for an inhomogeneous universe than for a 
homogeneous universe, we see that the potential barrier will be smaller for an
 inhomogeneous universe than for a homogeneous universe. Thus apparently an 
inhomogeneous universe will tunnel more easily through the classical barrier 
than the homogeneous universe. Since an inhomogeneous universe will have a 
larger Weyl tensor than an almost homogeneous one, we see that universes with 
large Weyl tensor tunnel more easily than those with a small Weyl tensor. 
  
If we look at the tunneling amplitude concerning effects from the $\Lambda$ 
term, it is evident that larger $\Lambda$ will yield a larger tunneling 
probability.  In the initial era inhomogeneities will increase the value of 
$\Pi$. We saw that  an inhomogeneous state will tunnel more easily through the
 potential barrier than a homogeneous state.  The largest tunneling 
probability amplitude thus occurs for universes with a large cosmological 
constant and large local inhomogeneities. From a classical point of view the 
value of $\Pi$ initially was large (but increasing thereafter), but as the 
universe entered the inflationary era the cosmological constant was relatively
 big so the value of $\Pi$ at the end of the inflationary era is relatively  
small. 

In the initial epoch the universe is not believed to be dust dominated. The 
dust does not exert any pressure and dust particles do therefore not interact 
with each other. A more probable matter content is matter which has internal 
pressure. Even though gravitation tends to make the space inhomogeneous, 
internal pressure from the matter will try to homogenize the space.  Since 
dust is the only  matter source in our model, the model only indicates the 
tendency for gravitation itself to create inhomogeneities. 

Since the universes tunnel into a deSitter-like state, the cosmological 
constant will rapidly dominate the evolution. The larger the cosmological 
constant the lower will the entity $\Pi$ be after the inflationary era ends.

\subsection{The Bianchi type I model}

The general solution of the WD-equation for the Bianchi type I models is (no 
scalar field):
\begin{equation}
\Psi(v,\beta)=\int d^2k\left[C(\vec{k})\psi_{\vec{k}}(v)\rho(\vec{k})
e^{i\vec{\beta}\cdot\vec{k}}\right]
\end{equation}
where $\psi_{\vec{k}}(v)=v^{-\frac{\zeta}{2}}W_{L,\mu}(2Hv)$ is a particular 
solution of the WD equation (with dust), $C(\vec{k})$ is a ``normalizing 
constant'', and $\rho(\vec{k})$ is a distribution function in momentum space. 
This distribution function satisfies the equation:
\[ \int d^2k |\rho(\vec{k})|^2=1 \]  

There is still a factoring problem in turning the classical entities to 
operators. Let us first investigate the expectation value of 
$C^{\alpha\beta\gamma\delta}C_{\alpha\beta\gamma\delta}$. Working in the small
 $v$ limit it does not seem possible to avoid the horrendous 
$\frac{1}{v^4}\left(\frac{\partial^2}{\partial \beta_{+}^2}+
\frac{\partial^2}{\partial \beta_{-}^2}\right)^2$ which will contribute with a
 term:
\[ \frac{\int d^2\beta \Psi^*\frac{1}{v^4}\left(\frac{\partial^2}{\partial 
\beta_{+}^2}+
\frac{\partial^2}{\partial \beta_{-}^2}\right)²\Psi}{\int d^2\beta \Psi^* \Psi}
 = \frac{1}{v^4}\frac{\int d^2k|C(\vec{k})|^2|\psi_{\vec{k}}(v)|^2
|\rho(\vec{k})|^2|\vec{k}|^4}{\int d^2 k|C(\vec{k})|^2|\psi_{\vec{k}}(v)|^2
|\rho(\vec{k})|^2}   
\]
Unless we have a delta-function distribution at $\vec{k}=0$; 
$\rho(\vec{k})=\delta^2(\vec{k})$, the contribution from this term to the 
Weyl Curvature invariant will diverge as $v^{-4}$. Thus, we have to conclude 
that, in the small $v$ limit the expectation value of 
$C^{\alpha \beta \gamma \delta}C_{\alpha \beta \gamma \delta}$ goes as:
\begin{equation}
\inl C^{\alpha \beta \gamma \delta}C_{\alpha \beta \gamma \delta}\inr\propto 
\frac{1}{v^4}
\end{equation} 
just as in the classical case. 
\par
Investigating the invariant $R^{\mu \nu}{R_{\mu \nu}}$, we notice that things 
are not so easy. The Ricci square also has a term which presumably would 
contribute with a $\frac{k^4}{v^4}$ term. However, looking at the classical 
expression we see that the Ricci square is independent of the anisotropy 
parameter $A$. This indicates that at the classical level all terms involving 
the anisotropy parameters, have to cancel exactly. This is not the case 
quantum mechanically. In the quantum case operators do not necessarily 
commute. Hence we may have contributions from  terms which classically would 
cancel each other. In other words, the fact that the classical vacuum has 
$R_{\mu\nu}=0$, does not mean that the quantum vacuum has $\hat{R}_{\mu\nu}=0$.

We assume that $(\xi_j)$ is a set of factor-ordering parameters which 
represents the ``true'' quantum mechanical system in such a way that  
$\xi_j=0$ represents the classical system. With this parameterization of the 
factor ordering we would expect the Ricci square expectation value to be:
\[ \inl R^{\mu \nu}{R_{\mu \nu}}\inr=4\Lambda^2+2\Lambda\frac{M}{v}+
\frac{M^2}{v^2}+\frac{f_j(v)}{v^4}\xi_j+{\mathcal{O}}(\xi_i^2) \]
where $f_i$ is some function of $v$ which has the property: 
$v\approx 0, \quad f_j(v)\approx constant$.
Thus for small $v$ and $\xi_j$ the expectation value would behave as
\[  \inl R^{\mu \nu}{R_{\mu \nu}}\inr \propto \frac{f_j(0)}{v^4}\xi_j \]

and
\begin{equation}
\frac{\inl C^{\alpha \beta \gamma \delta}C_{\alpha \beta \gamma \delta}\inr}
{\inl R^{\mu \nu}{R_{\mu \nu}}\inr} \propto\frac{1}{f_j(0)\xi_j}\cdot constant
\end{equation}
The Weyl square divided by the Ricci squared is in general finite as 
$v\longrightarrow 0$ for a quantum system. The actual value at $v=0$ is 
however strongly dependent on the factor-ordering. As the factor ordering 
parameters approach zero, the value will diverge. Quantum effects in the early
 epoch are essential for the behaviour of this entity near the initial 
singularity. As $v\longrightarrow 0$ we expect the quantum effect to be 
considerable, thus expecting that the factor-ordering parameters $\xi_j$ to be
 large.

As indicated already in section II the quantum mechanical expectation value of
 $\hat{P}$ will be lower and presumably finite in the initial singularity. 
Therefore the expectation value of $\hat{\Pi}$ is also presumed to be 
considerably lower in the initial stages than its classical counterpart. 

Comparing different tunneling amplitudes in the Bianchi type I model is 
difficult and more speculative because the Bianchi type I universe has no 
classically forbidden region for $\Lambda\geq 0$. This causes the lowest order
 WKB approximation to be purely oscillatory. The lowest order WKB wavefunction
 will therefore be approximately constant. In the paper \cite{sigBI} we did 
however construct under some assumptions a wavefunction which clearly peaked 
at small values of the anisotropy parameter. Thus these wavefunctions predicts
 universes that have a relatively low value of $\Pi$. 

\subsection{The effect of factor ordering}
As we saw the results obtained for the early part of the evolution of the 
universe is heavily dependent on the factor ordering. Even if we tried to 
capture some of the essentials of the factor ordering in some parameters, we 
did not actually know how we should order the operators. It appears however 
that the factor ordering we have used makes our quantities more well-behaved 
near the initial singularity. 

\section{Conclusion and Discussion}
When Penrose first suggested the Weyl curvature conjecture, the conjecture 
seemed reasonable because of  the tendency for gravitation to clump matter 
together and form inhomogeneities. Several investigations of the WCC showed 
that things where not so simple and some authors set the WCC in doubt. Most 
of these authors investigated local entities like $P^2$.  In a more recent 
paper by Rothman \cite{rothman} it appears as if Penrose even tried to 
withdraw the statement. But as we have shown in this paper there is no reason 
to withdraw the statement as long as we stick to non-local entities. In a 
early work by Husain\cite{Husain} it also appears that a non-local entity in a 
quantum inhomogeneous model, the Gowdy cosmology, is investigated. In that 
work a vacuum spacetime is considered where he looks at the behaviour of the 
Weyl tensor squared and his results are basically in agreement with ours. 
However entities like $\Pi$ and $P$ diverge for inhomogeneous vacuum 
spacetimes, hence for these spacetimes a more careful treatment is needed. 
Also 
Rothman and Anninos\cite{ra,rothman}  investigated a non-local 
entity\footnote{They used the volume of the phase space, similar to what we do
 for classical thermodynamical systems.} and their conclusion is similar to 
ours\footnote{When this paper was almost finished a paper by Pelvas and Lake 
\cite{pl} appeared, where they also investigated local entities like $P^2$. 
They also concluded that these entities cannot be a measure of gravitational 
entropy.}. Even if we cannot say that the Weyl tensor is directly linked to 
the gravitational entropy, we have shown that a certain non-local entity which
 is constructed from it has an entropic behaviour, and reflects the tendency 
of the gravitational field to produce inhomogeneities.

In the inflationary era the non-local entity $\Pi$ asymptotically evolved 
towards a constant. This is believed to be the correct behaviour in an 
inflationary era. The massive amount of vacuum energy tends to smooth out any 
inhomogeneities or anisotropies which is present from the pre-inflationary 
era. That this entity did not tend to zero (even though the co-moving volume 
expanded exponentially) could be interpreted to tell that $\Pi$ contained 
\emph{some} of the information from the pre-inflationary era. There were some 
seeds of the inhomogeneities left which could be the seeds needed to form 
galaxies as we see them today. 

\section*{Acknowledgments}
We would like to thank S. A. Johannessen for great support in the final stages
 of the work. We would also like to thank S. Hayward for his comments on this paper.


\begin{thebibliography}{99}
\bibitem{davies74}
P.C.W. Davies, {\it The Physics of Time Asymmetry}, Surrey University Press 
(1974).
\bibitem{davies83}
P.C.W. Davies, {\it Inflation and time asymmetry in the universe}, Nature, 
{\bf 301}, 398 (1983).
\bibitem{bekenstein}
J.D. Bekenstein, Phys. Rev. {\bf D7}, 2333 (1973).
\bibitem{hawking}
S.W. Hawking, {\it Particle creation by Black Holes}, Comm. math. Phys., 
{\bf 43}, 199, (1975).
\bibitem{penrose}
R. Penrose, in {\it General Relativity, an Einstein centenary survey}, 
eds. S.W. Hawking and W. Israel, Cambridge Univ. Press (1979).
\bibitem{wa}
J. Wainwright and P.J. Anderson, {\it Isotropic singularities and 
isotropization in a class of Bianchi Type-VI$_h$ cosmologies}, Gen. Rel. Grav.,
 {\bf 16}, 609 (1984).
\bibitem{ra}
T. Rothman and P. Anninos, {\it Hamitonian dynamics and the entropy of the 
gravitational field}, Phys. Lett. {\bf A224}, 227 (1997).
\bibitem{rothman}
T. Rothman, {\it A phase space approach to gravitational entropy}, 
Gen. Rel. Grav. {\bf 32}, 1185 (2000).
\bibitem{sigBI}
S. Hervik, {\it The Bianchi Type I minisuperspace model}, 
Class. Quantum Grav. {\bf 17}, 2765 (2000).
\bibitem{gw}
S.W. Goode and J. Wainwright, {\it Isotropic singularities in cosmological 
models}, Class. Quantum Grav., {\bf 2}, 99 (1984).
\bibitem{sigLT}
S. Hervik, {\it Quantum creation of an Inhomogeneous universe}, Class. Quantum Grav. {\bf 18}, 175 (2001)
\bibitem{miller}
B. Miller, {\it Negative-mass lagging cores of the Big Bang}, Astroph. J. 
{\bf 208}, 275 (1976).
\bibitem{Bon85}
W.B. Bonnor, {\it The gravitational arrow of time}, Phys. Lett. {\bf 112A}, 
26 (1985).
\bibitem{hayward}
S. Hayward, {\it Private Communications} (2000).
\bibitem{Zecca}
A. Zecca, {\it Il nuovo Cimento}, {\bf 106B}, 413 (1990).
\bibitem{Husain}
V. Husain, {\it Weyl tensor and gravitational entropy}, Phys. Rev. {\bf D38}, 3314 (1988)
\bibitem{pl}
N. Pelavas and K. Lake, {\it Measures of gravitational entropy: Self-similar 
spacetimes}, Phys. Rev. {D62}, 044009, (2000).
\end{thebibliography}
\end{document}